\documentclass[12pt,a4paper]{article}
 
\usepackage{epsfig} 
\usepackage{amssymb} 
\usepackage{amsmath}

\newcommand{\be}{\begin{equation}} 
\newcommand{\en}{\end{equation}} 
\newcommand{\bea}{\begin{eqnarray}} 
\newcommand{\ena}{\end{eqnarray}} 
 
\newcommand{\hbo}{\hbox to 1 true cm {\hfill } } 
 
\newcommand{\Tr}{\hbox{Tr}} 
 
\newcommand{\I}{{\rm i}} 
\newcommand{\E}{{\rm e}}

\newcommand{\re}[1]{~(\ref{#1})}

\newcommand{\hs}{\hspace{0.5cm}}

\begin{document} 
 
\vglue 1truecm 
 
\vbox{\hfill UNITU-THEP-006/2001 

} 
\vbox{\hfill  CERN-TH/2001-042
} 
   
\vfil 
\centerline{\large\bf Quantum diffusion of magnetic fields in a } 
\centerline{\large\bf numerical worldline approach }  
   
\bigskip 
\centerline{ Holger Gies$^{a,b,\ast}$ and Kurt Langfeld$^{b}$ } 
\vspace{1 true cm}  
\centerline{ $^a$ Theory Division, CERN, CH-1211 Geneva 23, Switzerland }  
 
\bigskip 
\centerline{and} 
\bigskip 
\centerline{ $^b$ Institut f\"ur Theoretische Physik, Universit\"at  
   T\"ubingen } 
\centerline{D--72076 T\"ubingen, Germany} 
\bigskip
$\text{}$

\bigskip
\centerline{ February 2001 }
   
\vfil 
\begin{abstract} 
  We propose a numerical technique for calculating effective actions 
  of electromagnetic backgrounds based on the worldline formalism. 
  As a conceptually simple example, we consider scalar 
  electrodynamics in three dimensions to one-loop order. Beyond the 
  constant-magnetic-field case, serving as a benchmark test, we 
  analyze the effective action of a step-function-like magnetic field 
  -- a configuration that is inaccessible to derivative expansions. 
  We observe magnetic-field diffusion, i.e., nonvanishing magnetic 
  action density at space points near the magnetic step where the 
  classical field vanishes. 
\end{abstract} 
 
\vfil 
\hrule width 5truecm 
\vskip .2truecm 
\begin{quote}  
${}^\ast$ {\small Emmy Noether fellow} 
 

keywords: worldline, effective action, Monte-Carlo simulation
\end{quote} 
\eject 
\section*{Introduction} 
The worldline formalism was invented by Feynman \cite{feynman1} 
simultaneously with modern relativistic second-quantized QED, but for
a long time it was used only occasionally for actual calculations. 
Eventually, the observation of a close relation between the worldline 
formalism and the infinite\--string\--tension limit of string path 
integrals triggered further developments of the worldline formalism 
for QCD \cite{berkos} and QED \cite{strassler}, particularly for gauge 
particle amplitudes. Certain computational advantages of this 
formalism were subsequently recognized and led to numerous 
applications. Among them, the progress achieved for QED amplitudes 
with all-order couplings to an external background field is 
particularly remarkable 
\cite{Schmidt:1993rk,Shaisultanov:1996tm,Reuter:1997zm}. The formalism 
works most elegantly for constant electromagnetic background fields 
and can be extended to a derivative expansion in the electromagnetic 
background \cite{Cangemi:1995by,Gusynin:1999bt}. For a comprehensive 
review of the worldline formalism in QED and beyond, see 
\cite{Schubert:2001he}.  
 
In the present work, we intend to demonstrate that the worldline 
formalism is moreover perfectly suited for numerical 
computations. This is because the path integral over closed loops in 
spacetime can be approximated by a finite ensemble of loops, which 
allows for a simple and fast evaluation of expectation values. The 
latter include observables depending on an {\em arbitrary} background 
field.  
 
We illustrate this proposal by means of a simple example: the 
one-loop effective action of Euclidean scalar QED in three dimensions.  
The computational task boils down to the calculation of the Wilson loop  
expectation value using a loop ensemble.  
 
As a first step, we verify the method in Sect.~\ref{constant} by 
considering the constant-magnetic-field case, which can also be solved 
analytically. This serves as a benchmark test for the concrete 
procedure that we propose. In a second step, the numerical method is 
applied to a magnetic field resembling a step function in one spatial 
direction (Sect.~\ref{diffusion}). This idealized field configuration 
represents the simplest configuration, which is inaccessible to the 
standard analytical method for inhomogeneous fields: the derivative 
expansion. As our main result, we observe diffusion of the magnetic 
field: the field takes influence on the region of vanishing  
background field by inducing a nonzero action density therein.  
 
Our conclusions are summarized in Sect.~\ref{conclusion}, where possible 
generalization and the road to further application of worldline 
numerics are sketched.

\section{The worldline approach \\ to functional determinants }  
\subsection{Setup}  
 
Our starting point is the unrenormalized Euclidean one-loop 
effective action of scalar QED in $D$ dimensions in worldline 
representation \cite{Schubert:2001he}, corresponding to the 
determinant of the gauge-covariant Klein--Gordon operator
\begin{equation} 
\Gamma^1[A]=\int\limits_{1/\Lambda^2}^\infty \frac{dT}{T}\, \E^{-m^2
  T}\, {\cal N}   
\int\limits_{x(T)=x(0)} {\cal D}x(\tau)\, \E^{-\int\limits_0^T d\tau 
  \left( \frac{\dot{x}^2}{4} +\I e\,\dot{x}\cdot
    A(x(\tau))\right)}, \label{1} 
\end{equation} 
where the superscript ``1'' indicates the one-loop level\footnote{In
  an abuse of nomenclature, we call the one-loop contribution to the
  total effective action also ``effective action'' in the following.
  The reader should always keep in mind that the Maxwell term (and all
  higher-order terms) must be added.}; a gauge-invariant UV
regularization at a scale $\Lambda$ has been performed at the lower
bound of the $T$ integral for the sake of definiteness.  In Eq.\re{1},
we encounter a path integral over closed loops in spacetime. Note that
there are no other constraints to the loops except differentiability
and closeness; in particular, they can be arbitrarily
self-intersecting and knotty. The normalization can be determined from
the zero-field limit,
\begin{equation} 
{\cal N}\int{\cal D}x(\tau)\, \E^{-\int\limits_0^T d\tau\, 
  \frac{\dot{x}^2}{4}} \stackrel{!}{=} \Tr\, \E^{\partial^2 T} =\int 
  \frac{d^Dp}{(2\pi)^D}\, \E^{-p^2T} =\frac{1}{(4\pi 
  T)^{D/2}}. \label{2} 
\end{equation} 
Solving Eq.\re{2} for ${\cal N}$ and inserting this into Eq.\re{1} 
leads us to the compact formula
\begin{equation} 
\Gamma^1[A]=\frac{1}{(4\pi)^{D/2}} \int d^Dx_0
\int\limits_{1/\Lambda^2}^\infty  
\frac{dT}{T^{(D/2)+1}}\, \E^{-m^2 T} \langle W[A]\rangle_x. \label{3} 
\end{equation} 
Here we have split off the integral over the zero-modes of the path 
integral, $\int d^Dx_0$, where $x_0$, the so-called loop center of mass,  
corresponds to the average position of the loop: 
$x_0^\mu:=(1/T)\int_0^Td\tau\,x^\mu(\tau)$. In Eq.\re{3}, we 
introduced the Wilson loop
\begin{equation} 
W[A(x)]=\E^{-\I e\int\limits_0^T d\tau\, 
  \dot{x}(\tau) \cdot A(x(\tau))} \equiv \E^{-\I e\oint dx\cdot A(x)},
  \label{4}   
\end{equation} 
and $\langle(\dots)\rangle_x$ denotes the expectation value of (\dots) 
evaluated over an ensemble of $x$ loops; the loops are centered upon 
a common average position $x_0$ (``center of mass'') and are 
distributed according to the Gaussian weight $\exp[-\int_0^T 
  d\tau\, \frac{\dot{x}^2}{4}]$.  
 
The following $\tau$-integral substitution, $\tau=:T t$, is of crucial 
importance for the numerical realization of the path integral; it 
suggests introducing {\em unit loops} $y$, 
\begin{equation} 
y(t):=\frac{1}{\sqrt{T}}\, x(T t), \quad t\in [0,1], \label{5} 
\end{equation} 
which are parameterized with a unit propertime $t$. The remaining 
integrals can be rewritten accordingly: 
\begin{equation} 
\int\limits_0^T\!d\tau\, \frac{\dot{x}^2(\tau)}{4}  = 
   \int\limits_0^1\!dt\, \frac{\dot{y}^2(t)}{4}, \quad  \!
\int\limits_0^T\!d\tau\,\dot{x}(\tau)\cdot A(x(\tau))
   =\int\limits_0^1\! dt\,  \dot{y}(t)\cdot [\sqrt{T}
   A(\sqrt{T}y(t))]. \label{6}  
\end{equation} 
The important advantage is constituted by the fact that the expectation 
value of $W[A]$ can now be evaluated over the unit-loop ensemble $y$, 
\begin{equation} 
\langle W[A(x)]\rangle_x \equiv \langle 
W[\sqrt{T}A(\sqrt{T}y)]\rangle_y,\label{7} 
\end{equation} 
where the exterior $T$-propertime dependence occurs only as a scaling 
factor of the gauge field and its argument. In other words, while 
approximating the loop path integral by a finite ensemble of loops, 
it suffices to have one single unit-loop ensemble at our disposal; we 
do not have to generate a new loop ensemble whenever we go over to a 
new value of $T$.  
 
Inserting Eq.\re{7} into Eq.\re{3}, we arrive at the final formula 
(dropping the subscript of $x_0$):
\begin{equation} 
\Gamma^1[A]=\frac{1}{(4\pi)^{D/2}} \int d^Dx
\int\limits_{1/\Lambda^2}^\infty  
\frac{dT}{T^{(D/2)+1}}\, \E^{-m^2 T} \biggl\langle 
W \Bigl[\sqrt{T}A(\sqrt{T}y) \Bigr] \biggr\rangle_y \; + \;  
{\rm c.t.}, \label{8}  
\end{equation} 
where we have formally added counter-terms
(c.t.) which correspond to a renormalization of 
the physical parameters. The details of the c.t.'s depend on the 
dimension, but not on the type of the background field, and will 
be discussed in the following. 
 
\subsection{Renormalization}  
 
For an accurate evaluation of the effective action of Eq.\re{8},  
an {\em analytic} calculation of the counter-terms is inevitable in 
order to obtain accurate results by the {\em numerical} procedure.  
In the following, we will confine ourselves to the important cases  
of $D=3$ and $D=4$ dimensions at the one-loop level.  
 
Let us first discuss the analytical treatment: while for $D=3$ the 
effective action is rendered finite by dropping a field-independent 
constant, the complete effective action in the 4-dimensional case is 
renormalized in such a way that the quadratic term in the field 
strength is given by 
\begin{equation} 
\Gamma _{\rm quadr.} \; = \; \frac{1}{4 g_R^2} \int d^4x \;  
F_{\mu \nu } (x) \, F_{\mu \nu } (x) \; ,  
\label{eq:9}  
\end{equation} 
where $F_{\mu \nu } (x)$ is the field strength tensor, and $g_R$ 
represents the renormalized coupling. In practice, this is done by 
subtracting the (infinite) quadratic term of the one-loop contribution 
$\Gamma^1$ and absorbing it into the bare Maxwell term. 
 
These UV divergencies of the effective action emerge from the lower 
bound of the propertime integral in Eq.\re{8}, i.e., $T \rightarrow 0$. 
For small values of the propertime, the Wilson loop expectation value 
can be calculated exactly (e.g., using heat-kernel techniques): 
\begin{equation}  
\biggl\langle W[\sqrt{T}A(\sqrt{T}y)] \biggr\rangle_y \; = \;  
1 \; - \; \frac{1}{12} T^2 \, F_{\mu \nu }[A] (x) \, F_{\mu \nu }[A] (x)  
\; + \; {\cal O} (T^4) \; ,  
\label{10}  
\end{equation}  
where $x$ is the loop center of mass. The key observation is that,  
even for $D=4$, the terms of order $T^4$ are UV-finite upon the  
propertime integration in Eq.\re{8}. Hence, the completely renormalized  
effective action, which is suitable for numerical simulations,  
is  
\begin{eqnarray}   
\Gamma^1[A] = 
\frac{1}{(4\pi)^{D/2}} \int d^Dx \; \int _0^\infty 
\frac{dT}{T^{(D/2)+1}}\, \E^{-m^2 T}  
\biggl[  &&\!\!\!\!\!\!\!\! 
\Bigl\langle W[\sqrt{T}A(\sqrt{T}y)] - 1 \Bigr\rangle_y  
\label{11} \\  
&&\!\!\!\!\!\!\!+\frac{1}{12} \; T^2 \, F_{\mu \nu }[A] (x) \, F_{\mu 
  \nu }[A] (x)  \biggr]\nonumber\\ 
+c_D \; \int d^Dx \;  F_{\mu \nu }[A] (x) \, F_{\mu \nu }[A] (x) \; ,  
\nonumber  
\end{eqnarray}  
where  
\begin{equation}  
c_3 \; = \; -\frac{1}{96\pi} \frac{1}{m}+{\cal O}(1/\Lambda)\; ,  
\quad c_4 \; =-\frac{1}{12} \frac{1}{16\pi^2}\left( \ln 
  \frac{\Lambda^2}{m^2} -C\right)+{\cal O}(m^2/\Lambda^2).  
\label{12}  
\end{equation}  
Here $\Lambda$ denotes the UV cutoff, and $C$ is Euler's number. In 
the $D=4$ case, the term $\sim c_4$ will be absorbed into the bare 
Maxwell term, defining the running of the coupling, and the cutoff can
subsequently be sent to infinity. In $D=3$, the 
theory is super-renormalizable and the term $\sim c_3$ represents only 
a finite shift of the coupling, introducing no running. In the first
two lines of Eq.\re{11}, we have already sent the cutoff $\Lambda$ two
infinity, leaving us with a perfectly finite expression.  Note that in 
the case $D=3$, the term of order $T^2$ in Eq.\re{10} corresponds to a 
singularity $1/\sqrt{T}$, which is integrable. Hence, the subtraction 
of this term from the propertime integral is not mandatory. However, 
aiming at worldline numerics, the perfect control over the behavior of 
the integrand at small propertime $T$ is required in order to augment 
the accuracy of the numerical evaluation of the propertime integral. 
 
Now, the numerical renormalization is similar in spirit to the 
analytical one, but is complicated by a further problem: evaluating 
$\langle W\rangle$ with the aid of the loop ensemble does not produce 
the small-$T$ behavior of Eq.\re{10} {\em exactly}, but, of course, 
only within the numerical accuracy. Unfortunately, even the smallest 
deviation from Eq.\re{10} will lead to huge errors, if we naively plug 
such a result into Eq.\re{12}; this is because of the factors of $T$ 
in the denominator of the integrand for $T\to 0$.  
 
Our solution to this problem is to fit the numerical result for 
$\langle W\rangle$ to a polynomial in $T$ in the vicinity of $T=0$, 
employing Eq.\re{10} as a constraint for the first coefficients. Of 
course, such a fit is completely justified because of our exact 
knowledge of $\langle W\rangle$ for small $T$. This fit not only 
represents the renormalization procedure, solving the UV problems, but 
also facilitates a more precise estimate of the error bars (see 
below). Finally, employing this fitting procedure only close to $T=0$, 
the infrared behavior ($T\to \infty$) of the integrand remains 
untouched, and our approach is immediately applicable, also in the case 
$m=0$.  
 
Both renormalization procedures, the analytical as well as the 
numerical, generalize straightforwardly to higher dimensions; here, 
either additional subtractions from the integrand (for the analytical 
case), or polynomial fits with higher-order constraints (for the 
numerical case) are required: for example, increasing the dimension 
by 2 requires one more subtraction/constraint in the small-$T$ 
behavior of the integrand. 
 
Finally, it should not be concealed that these procedures become 
increasingly difficult to higher order in perturbation theory for 
massive theories in $D\geq4$. Then, a mass renormalization is also 
necessary, requiring careful analyses of the UV behavior of double 
propertime integrals \cite{Ritus:1975cf}.

\subsection{Numerical simulation }  
 
Now, the route to the effective action is clear: 
\begin{enumerate} 
\item[(1)] generate a unit loop ensemble distributed according to the 
  weight \break $\exp[ -\int_0^1dt \dot{y}^2/4]$, e.g., employing the 
  technique of normal (Gaussian) deviates; 
\item[(2)] compute the integrand for arbitrary values of $T$ (and 
  $x$); this involves the evaluation of the Wilson loop expectation 
  value for a given background gauge field;  
\item[(3)] perform the renormalization procedure and add the c.t.'s to
  the integrand; 
\item[(4)] integrate over the propertime $T$ in order to obtain the 
Lagrangian, and also over $x$ for the action. 
\end{enumerate} 
 
From a general viewpoint, there are two sources of systematical error 
which are introduced by reducing the degrees of freedom from an 
infinite to a finite amount: first, the loop path integral has to be 
approximated by a finite number of loops; second, the propertime $t$ 
of each loop has to be discretized. Contrary to this, the spacetime 
does not require discretization, i.e., the loop ensemble is generated 
in the continuum. 
 
Of course, the various steps can be carried out using different
numerical methods; let us outline our choice of tools in detail. For
this purpose, let $N_l$ denote the number of loops which are used to
estimate the Wilson loop expectation value in Eq.\re{7}, and $n_l$ the
number of space points which are employed to specify a particular
loop.  The points of a particular loop $y_i, i=1 \ldots n_l$, are
generated by a standard heat bath algorithm, where the boundary
conditions $y_1=0$, $ y_{nl} =0 $ are enforced.  After a proper
thermalization, all coordinates $y_i$ are shifted equally in order to
implement the center of mass condition $\langle y_i \rangle =x_0$.
This procedure is then repeated $N_l$ times to generate the loop
ensemble.
 
Approximating $\langle W \rangle $ of Eq.~(\ref{7}) by an average over 
a finite number of loops, the standard deviation provides an estimate 
of the statistical error. Approximating the loops by the finite number 
$n_l$ of space points results in a systematic error that can be 
estimated by repeating the calculation for several values $n_l$. The 
number $n_l$ should be chosen large enough to reduce this 
systematic error to well below the statistical one.  It will turn out 
that the choice $N_L=1000$ and $n_l=100$ yields results, for the 
applications below, which are accurate at the per cent level. We stress, 
however, that $\approx 20000$ (dummy) heat bath steps are required to 
properly thermalize the loop ensemble.

\section{Benchmark test:\\ constant magnetic background field} 
\label{constant} 
 
Since analytic results are available for the case of a constant 
magnetic background field $B$, we will investigate in this section the 
efficiency of our numerical loop approach to the scalar functional 
determinant.\footnote{Incidentally, another exactly solvable system is 
  known \cite{Cangemi:1995ee} with a background field resembling a 
  solitonic profile in one dimension.} For simplicity, we consider
three Euclidean spacetime dimensions, $D=3$. Up to a field-independent 
constant, the one-loop effective-action density (Lagrangian) ${\cal 
  L}^1$ for scalar QED in $D=3$ is given by  
\cite{Schubert:2001he,Schwinger:1951nm} 
\begin{equation}  
{\cal L}^1\equiv\Gamma^1 (B) / V_3 \; = \; \frac{B^{3/2}}{(4 \pi )^{3/2}} \;  
g \left( \frac{m^2}{B} \right) \; ,  
\label{13}  
\end{equation}  
where $V_3$ denotes the volume and  
\begin{equation}  
g(z) \; = \; \int _0^\infty \frac{ dT }{ \sqrt{T} } \; I(T,z) \; ,  
\hbox to 1cm {\hfill }  
I(T,z) = \frac{1}{T^2} \biggl[  
\frac{T }{\sinh T} \, - \, 1 \biggr] \; \exp \{ -zT \}  
\; .  \label{14}  
\end{equation}  
 
\begin{figure}[t] 
\centerline{  
\epsfxsize=7cm 
\epsffile{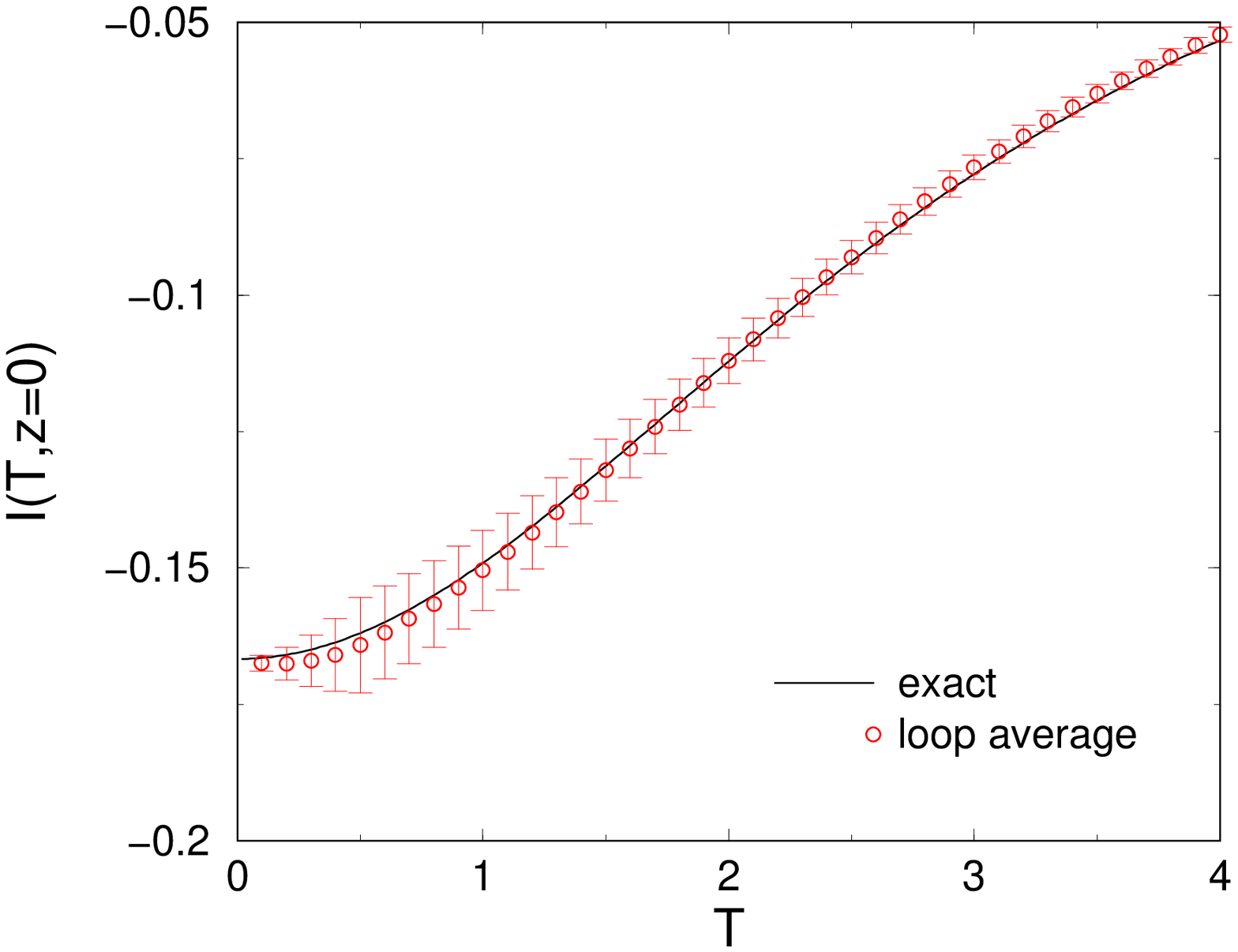} 
\epsfxsize=7cm 
\epsffile{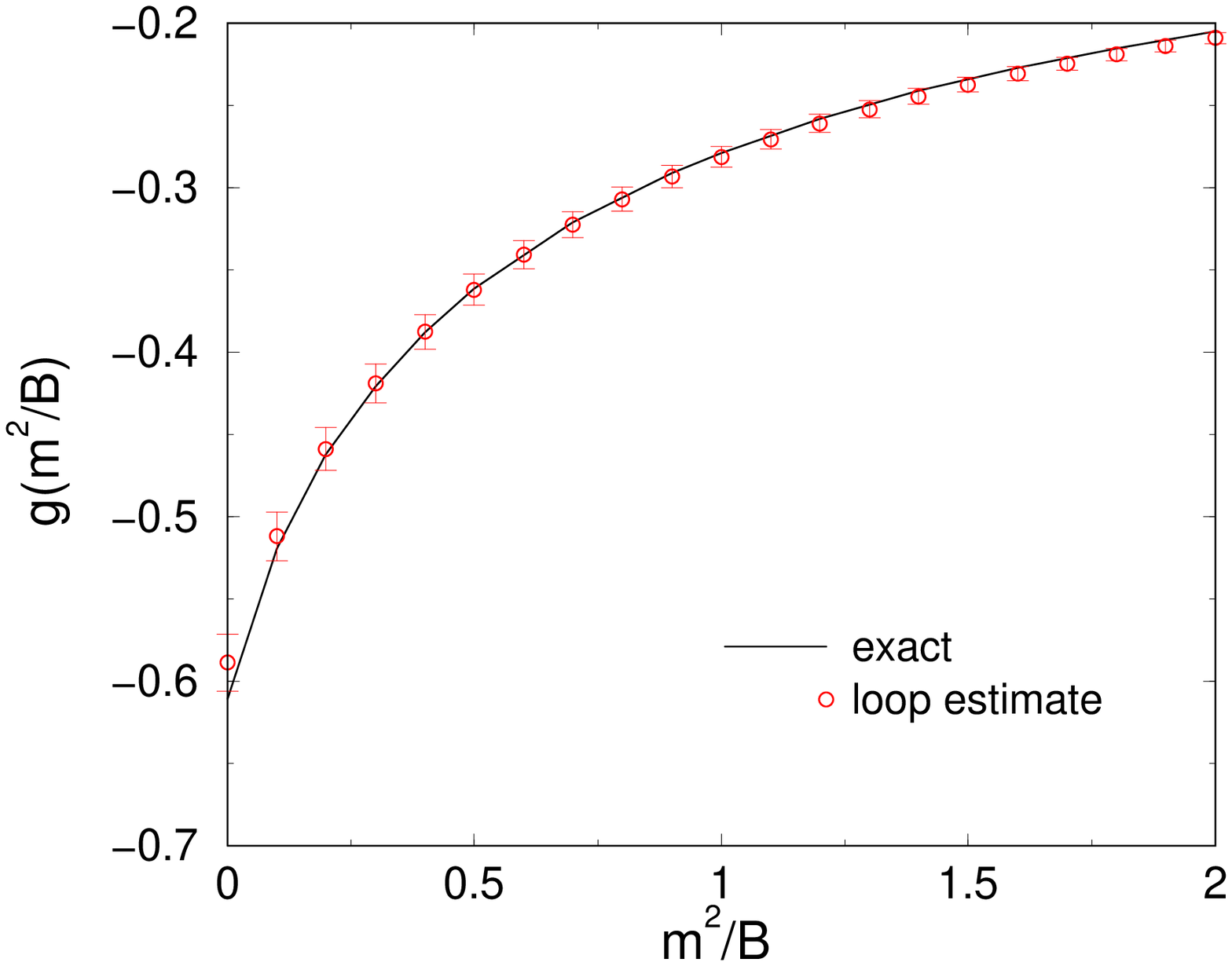} 
} 
\caption{\small Propertime integrand $I(T,z=0)$ (left panel) and integral 
  $g(m^2/B)$ (right panel) of the one-loop effective action for the 
  case of a constant magnetic background field. The analytically known 
  exact results (solid lines) are compared with the numerical findings 
  (circles with error bars). } 
\label{fig:1}  
\end{figure}

The exact integrand $I(T,z)$ in Eq.\re{14} is compared with our loop
estimate in Fig.~\ref{fig:1} for the case $m^2=0$, i.e., $z=0$.  The
numerical estimate for $g(m^2/B)$ is also shown. The agreement between
the two curves is satisfactory and the exact results for $I(T,z)$ as
well as for $g(m^2/B)$ lie well within the error bars produced by the
numerics.
 
As mentioned above, the error bars correspond to the statistical error 
of the ensemble average. Regarding the quantity $\langle W\rangle$ 
with its dependence on $T$, the original error bars are rather 
independent of $T$; but multiplying $\langle W\rangle$ by a 
$T$-dependent function (cf. Eq.\re{8}) causes a modulation of the error 
bars. In particular for $T\to 0$, the $1/\sqrt{T}$ singularity leads 
to an unbounded enhancement of the error bars for small $T$ in the 
function $I(T,x)$ (see left panel of Fig.~\ref{fig:1}). Fortunately, 
the integrand in this regime is known {\em exactly} as given in 
Eq.\re{10} in the form of an asymptotic series.  As mentioned above 
for the numerical renormalization procedure, this information can be 
used for a constraint fit of the numerical result to a polynomial in 
$T$ around $T=0$, keeping the first terms corresponding to Eq.\re{10} 
fixed.  Then, the error bars of the ensemble average translate into 
errors for the higher coefficients of the polynomial. This polynomial 
is then matched to the pure numerical result at that value of $T$ 
where the error bars of the two results are comparable. The final result 
is visible on the left panel of Fig.~\ref{fig:1}. 

It should be noted, that the error bars in Fig.~\ref{fig:1} (and the
following figures) are highly correlated from point to point, since
the same loop ensemble has been used for the evaluation of each point.
This correlation can, of course, easily be reduced by updating the
loop ensemble with a few heat bath steps inbetween at the expense of
computational time.

\section{Magnetic-field diffusion} 
\label{diffusion} 
 
In this section, we study the one-loop effective action of
Eq.\re{11} for the case of a magnetic background field, resembling
a step function in space. In particular, we consider a time-like
constant background field $B$, i.e., a field which is independent of
the third coordinate called Euclidean time.  We choose the $B$ field,
being a (pseudo-)scalar over the spatial $xy$ plane, as
\begin{equation}  
B(x,y) \; = \; - \theta (x) \, B_0   \; ,  
\quad 
\vec{A}(x,y) \; = \; \theta (x) \; \frac{1}{2} \, (y,-x) \; B_0 \; ,  
\label{eq:d1}  
\end{equation}  
where $\theta (x)$ is the step function, i.e., 
$$ 
\theta (x) \; = \; 1 \; , \; \; \forall x \ge 0 \; ,  
\hbox to 1cm {\hfill }  
\theta (x) \; = \; 0 \; , \; \; \forall x < 0 \; .  
$$  
 
Note that because of the sudden variation of the background field at
the step, the effective action cannot be obtained within a derivative
expansion. By contrast, such a discontinuity represents no obstacle
for the numerical worldline approach proposed in this work.
Discontinuities do not induce (artificial) singularities, but are
smoothly controlled by the properties of the loop ensemble. This
ensemble resembles a cloud, exhibiting finite extension and slowly
varying density, and being centered at some particular spacetime point
$x_0$ which is the center of mass of each loop in the cloud. While
running with this $x_0$ towards and across the step, that part of the
volume of the loop cloud which ``feels'' the magnetic field increases
smoothly; therefore, the effective-action density (effective
Lagrangian) ${\cal L}^1$, being the propertime integrated information
of what the loop cloud measures, will be smooth, too.
 
\begin{figure}[t] 
\centerline{  
\epsfxsize=7cm 
\epsffile{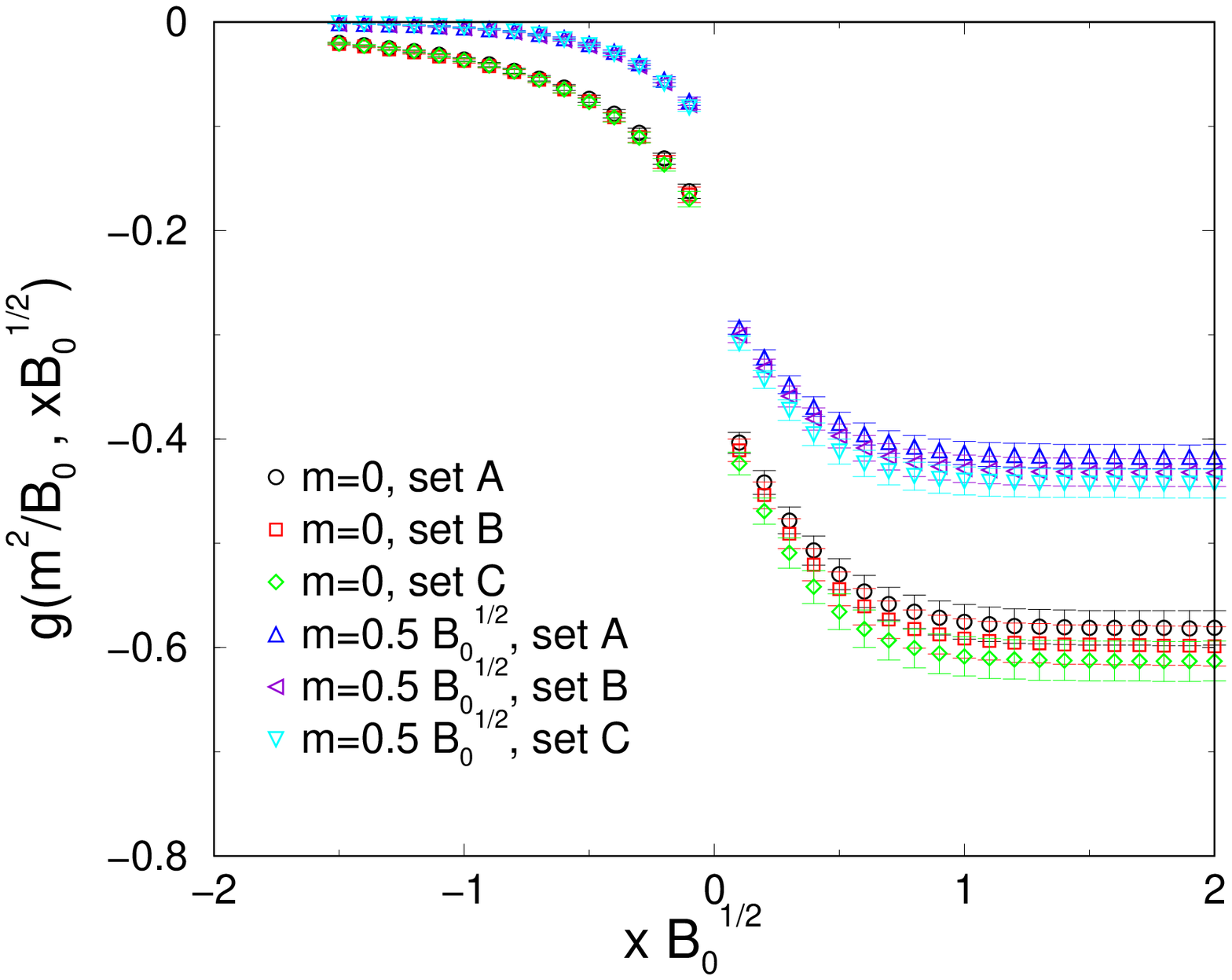} 
\epsfxsize=7cm 
\epsffile{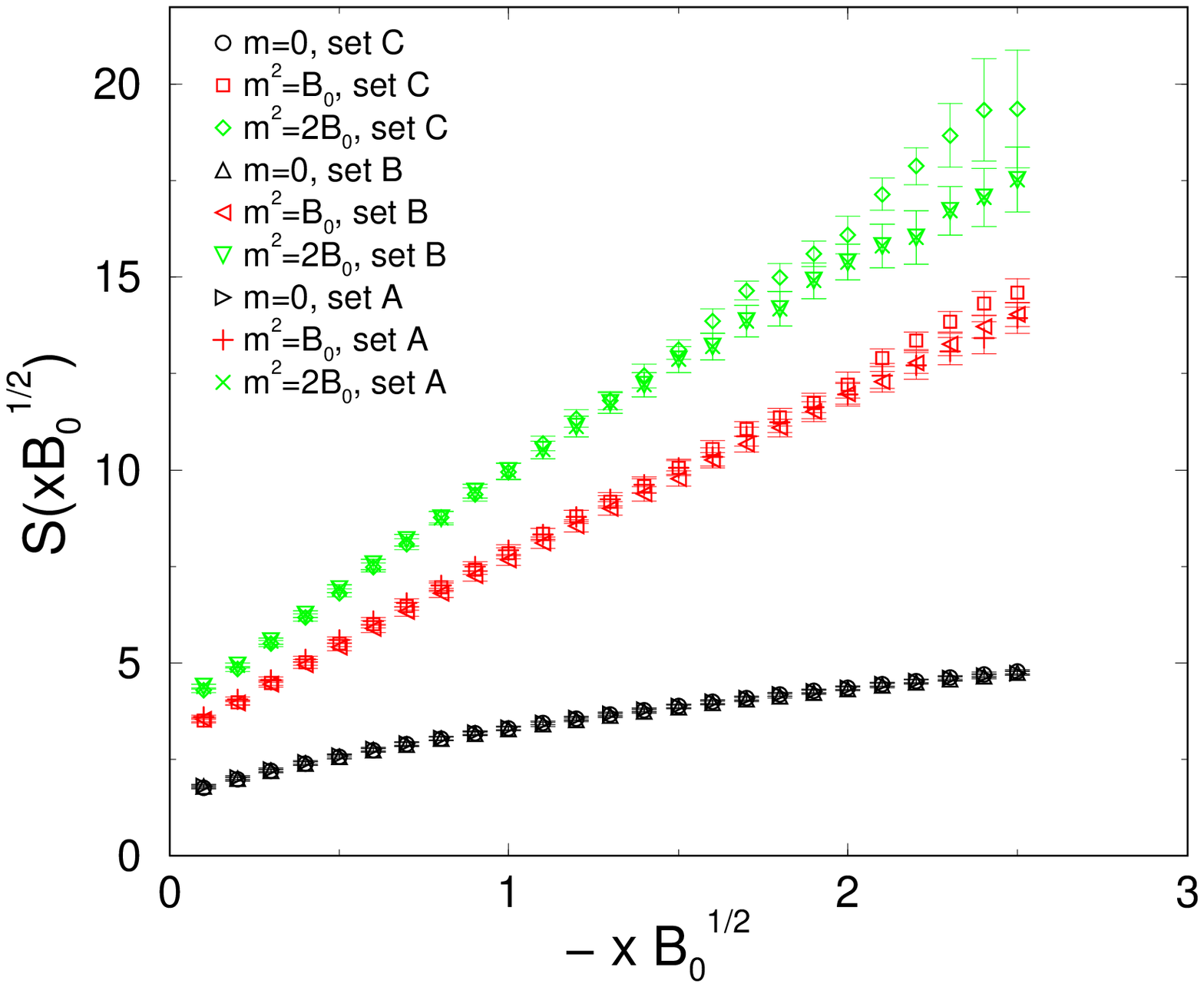} 
} 
\caption{\small Effective-action density in the vicinity of the 
  magnetic step ($x=0$). On the left panel, the diffusion profile is 
  visualized. The logarithmic plot on the right panel reveals the 
  exponential nature of the diffusion depth. } 
\label{fig:2}  
\end{figure}

In order to numerically evaluate the effective action for the case of
Eq.\re{eq:d1}, we employ the loop approach outlined in the previous
section. This approach provides for statistical error bars which allow to
estimate whether the number of loop ensembles is large enough. In
order to study the systematic errors, such as the limited number of
spacetime points $n_l$ specifying a loop and the limited number $n_T$ of
thermalization sweeps, we shall employ three sets of loop ensembles
(see Table 1). Defining
\begin{table}[b]
\caption{ Simulation parameters } 
\label{tab:1}
\begin{center}
\begin{tabular}{ccc} \hline\hline
          & \hs $n_l$  & \hs $n_T$   \\ 
set A \hs & \hs 100 & \hs 150000     \\ 
set B \hs & \hs 75  & \hs 50000       \\ 
set C \hs & \hs 50  & \hs 50000       \\ \hline\hline
\end{tabular}
\end{center}
\end{table}
\begin{equation}
g\left( \frac{m^2}{B_0^2},xB_0^{1/2}\right):=
 \frac{(4\pi)^{3/2}}{B_0^{3/2}}\, {\cal L}^1(\vec{x}=(x,0,0))
 \label{gdef}
\end{equation}
in analogy to Eq.\re{13}, the final numerical result for the effective
action is shown in Fig.~\ref{fig:2}, left panel.  As expected, the
effective-action density is nonzero even in the region $x<0$ where the
background field $B(\vec{x})$ vanishes. For increasing mass, the
contributions from large propertime $T$ are exponentially suppressed
in the integrand. Since the propertime controls the size of the loop
cloud, the large loop clouds contribute less to the action density
when the mass is large.  Hence, the effective-action density for an
increasing mass becomes reduced.  (From an alternative viewpoint, the
limit of large mass and small field are identical for dimensional
reasons.)

We find that the effective action decreases exponentially in the 
``forbidden'' region, which is depicted in Fig.~\ref{fig:2}, right
panel; here we defined the quantity 
\begin{equation}  
-g\left(\frac{m^2}{B_0},xB_0^{1/2}\right)\; =: \; \exp \left[ - S \left(x 
    B_0^{1/2} \right) \right],
\label{eq:d2}  
\end{equation}  
and plotted $S(x B_0^{1/2})$ for several values of the scalar mass $m$.
This phenomenon of obtaining a nonzero effective-action density even
in a region where the background field vanishes may be called {\it 
  quantum diffusion} of the magnetic field. 

\begin{figure}[t] 
\centerline{  
\epsfxsize=12cm 
\epsffile{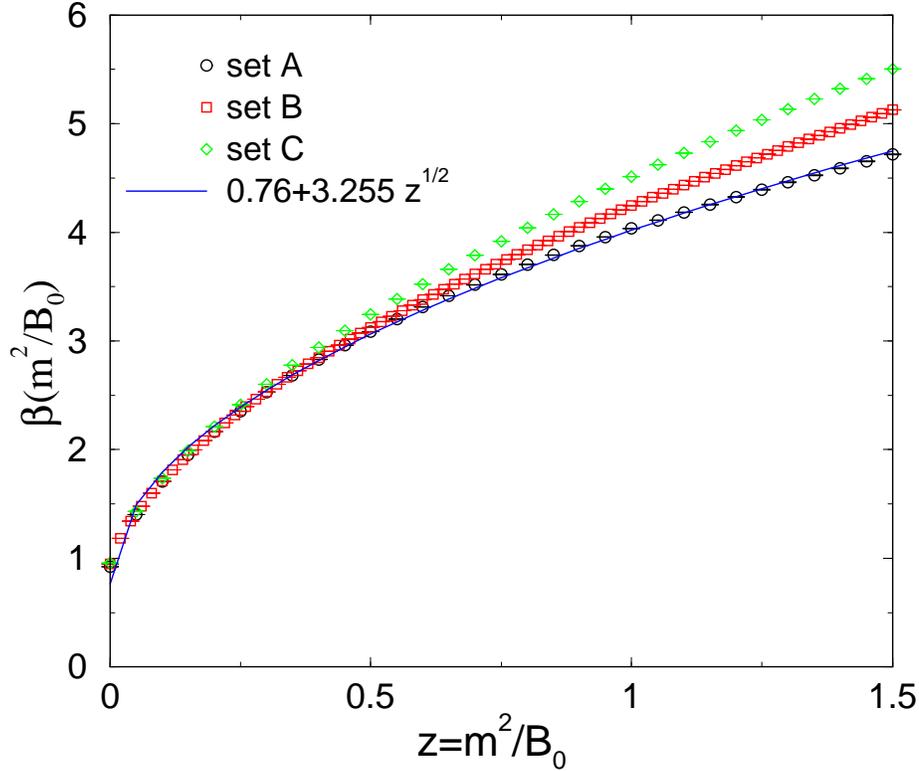} 
} 
\caption{\small Inverse diffusion depth $\beta$ versus
  mass-to-magnetic-field ratio. The plot symbols depict the numerical
  results, whereas the solid line represents the fit of
  Eq.\re{eq:fit2} adjusted to the optimized loop ensemble A.} 
\label{fig:3}  
\end{figure}

Large values of $S$ imply that the effective-action density $\sim
g\left({m^2}/{B_0},xB_0^{1/2}\right)$ is small due to
cancellations. In this case, a sufficiently large number of loop
points $n_l$ and of thermalization sweeps is requested. The results
indicate that the function $S(xB_0^{1/2})$ can be fitted for
$x\sqrt{B_0}$ by the ansatz
\begin{equation}
S(y)=\alpha+\beta\, y, \quad y=xB_0^{1/2}, \label{fit1}
\end{equation}
where the quantities $\alpha$ and $\beta$ depend on $m^2/B_0$.  The
diffusion depth $l$ of the magnetic field can be defined as the
inverse of $\beta$, i.e., $l=1/(\beta \sqrt{B_0})$. An estimate of
this parameter has been plotted versus the mass $m$ in
Fig.~\ref{fig:3}. As expected, the diffusion coefficient $\beta $
rises with increasing mass scale $m$.  The function
\begin{equation}
\beta (m^2/B_0) \; = \; 0.7627 \, + \, 3.255 \, \left( \frac{m^2}{B_0} \right) 
^{1/2} 
\label{eq:fit2} 
\end{equation}  
nicely fits the high quality numerical data set A (see figure
\ref{fig:3}). 

In order to get an understanding of these numbers, let
us perform the following heuristic consideration: at a first glance,
one might expect that the magnetic diffusion obeys the simple law
$\sim \E^{-\text{const.} \cdot m x}$, since the mass seems to be the
only scale in the field-free region of space; this would correspond to
$\beta=\text{const.} \cdot (m^2/B_0)^{0.5}$. However, if this simple
law were true, the massless limit $m=0$ would be obscure, since then
the magnetic field would diffuse into the field-free region without
any damping. Hence, there must be an additional dependence of the
exponent on the magnetic field in order to account for a reasonable
massless limit. Moreover, the mass is indeed not the only scale in the
field-free region, because the loop cloud is a nonlocal object that
always ``feels'' the strength of the magnetic field.  In fact, it is
the constant first term in $\beta$ in Eq.\re{eq:fit2} that exactly
accounts for this dependence of the magnetic diffusion on the strength
of the magnetic field. Only for larger masses (or weaker fields),
$m^2\simeq B_0/4$, the intuitively expected diffusion law of the form
$\sim \E^{-\text{const.}\cdot m x}$ begins to dominate; in this
regime, we find 
\begin{equation}
l \; \approx \; 0.31 / m \; .
\label{eq:diff} 
\end{equation} 
In the present work, the precision of the numerics for the action
density in the field-free region restricts the investigation to mass
values of $m^2 < 1.5 B_0$; beyond this, the strong exponential
decrease beyond the step prohibits a reliable analysis of the
diffusion depth.

\section{Conclusion and Outlook} 
\label{conclusion} 
      
Beyond any particular result of the present work, we would like to
remark in the first place that our approach to the worldline formula
(see Eq.\re{8}) for the one-loop effective action offers a vivid
picture of the quantum world. Consider a spacetime point $x$ at a
propertime $T$; then, the loop ensemble is centered upon this point
$x$ and resembles a loop cloud with Gaussian ``density'' and
``spread''. Increasing or lowering the propertime $T$ corresponds to
bloating or scaling down the loop cloud or, alternatively, zooming out
of or into the microscopic world. The effective-action density at each
point $x$ finally receives contributions from every point of the loop
cloud according to its Gaussian weight (times the mass term and other
factors) and averaged over the propertime. This gives rise to the
inherent nonlocality and nonlinearity of the effective action, because
every point $x$ is influenced by the field of any other point in
spacetime experienced by the loop cloud.
 
In particular, we considered the one-loop contribution to the 
effective action in scalar electrodynamics in three spacetime 
dimensions; in the beginning, we were able to reproduce the 
analytically well-known case of a constant magnetic background field, 
serving as a testing ground for our numerical procedures including 
renormalization to one-loop order. Incidentally, the zero-mass limit 
(or, alternatively, the ultra-strong magnetic field limit) is also 
covered by our approach without additional difficulties. 
 
We furthermore tackled the problem of a step-function-like magnetic
background field, illustrating the stability of our approach also for
discontinuous field configurations. For this case, we observed a
diffusion of the magnetic field, i.e., nonzero action density even at
a distance from the magnetic field. This phenomenon is obviously
nonlocal, since an expansion around a point of zero background field
gives a zero result to any finite order. But the diffusion phenomenon
appears to be also nonperturbative in the same sense as the Schwinger
mechanism of pair production \cite{Schwinger:1951nm}, at least for not
too large values of the mass; this is suggested by the functional form
of the diffusion for small mass: $exp(-S(xB_0^{1/2}))\sim \exp(-0.7627
\cdot \sqrt{B_0} x)$; this result cannot be expanded in terms of the
coupling constant, being rescaled in the field (only an expansion in
terms of the square root of the coupling constant is possible). 
 
Perhaps the main advantage of the numerical worldline approach to 
functional determinants is that no a-priori information, such as 
certain symmetries of the background field or a suitably chosen set of 
base functions, has to be exploited; if the loop ensemble has been 
thermalized properly, any background field can immediately be plugged 
into the algorithm. 
 
One drawback of the numerical approach lies in the fact that it 
applies only to Euclidean quantum field theory; this is because the 
action governing the distribution of the loops must be positive.  
 
The present paper paves the way to further generalizations such as  
fer\-mio\-nic functional determinants; our approach could  
facilitate a systematic numerical investigation of these  
determinants. Results could be compared with several results known from 
analytical considerations (see, e.g., \cite{Fry:1997an}). Although the 
treatment of fermions in the worldline formalism can be most elegantly 
formulated via Grassmann representations (see, e.g., 
\cite{D'Hoker:1996ax}), a numerical approach has to rely on a bosonic 
representation of the path integrals \cite{feynman1,Barut:1989ud}. In 
practice, this means that the Dirac algebraic elements in the 
worldline action are accompanied by a path-ordering prescription. 
Similar complications occur for nonabelian gauge fields and 
color-charged fermions or scalars. Whereas such path ordering is 
difficult to deal with in analytical approaches, a numerical 
evaluation can immediately take care of such a prescription. This is 
because the loops are discretized in the propertime parameter anyway, 
consisting of $n_l$ ``links''; the path ordering then is nothing but a 
simple (matrix-)multiplication of these links. 
 
Recently~\cite{la00}, a new approach was designed to improve the  
still un-pleasant situation~\cite{bar99} when lattice QCD is studied at  
finite baryon densities: the basic idea to circumvent  
the problems~\cite{ste96} with lattice fermions is a calculation of  
the {\it continuum } fermion determinant for arbitrary entries of the 
gluon field, which subsequently is the subject of a lattice discretization.  
This program was successfully applied to the case of heavy  
quarks~\cite{la00}. The formulation of the present paper stirs the hope  
that the concept of~\cite{la00} might be extended to the realistic case  
of light quarks.  
 
Further possible and straightforward applications can be found in the
context of estimating quantum energies to solitonic fields
(see~\cite{gra00}), or in the context of thermal field theory or
Casimir vacua (for analytical worldline approaches, see, e.g.,
\cite{McKeon:1993if}). The latter cases are connected with a
compactification of the Euclidean spacetime manifold; these
topological modifications can be imposed directly on the loop ensemble
in our approach. For example, compactifying the Euclidean time
direction at finite temperature will result in closed loops that wind
several times around the time-like cylinder. Work in these directions
is in progress.
  
Finally, it is obvious that the step-like magnetic field is not
physical at all in a strict sense, because such a sharp drop-off
cannot be produced by real laboratory magnets; nevertheless, the
step-like field can be regarded as a limiting case of a real physical
situation, being useful for an estimate of possible effects caused by
rapid variations of a magnetic field. For instance, as a matter of
principle, magnetic diffusion can be included in the discussion of the
Aharonov-Bohm effect: if the propagating particle is no longer
considered as a quantum mechanical point particle, but as a quantum
field, it will be affected by the quantum diffusion of the solenoid's
magnetic field; but a measurable effect can only arise, if the
particle's distance from the solenoid is of the order of a few Compton
wavelengths. 

Beyond that, we would like to mention that our results for the
step-like magnetic field are of some significance for the experiments
being currently performed at PVLAS (Legnaro) and BMV (Toulouse)
\cite{Exp}, in which is measured the optical birefringence of the
quantum vacuum exposed to a magnetic field. For example in the PVLAS
experiment, the polarization axis of a laser beam is affected by a
(6--9 T)-magnetic field with a diameter of 1 m; the magnetic field
drops off over a distance of 10 cm, and the question arises as to
whether this drop-off will influence the rotation of the polarization
axis as predicted by a constant-field QED calculation.
 
The refractive indices of the modified vacuum are proportional to the 
energy density induced by the magnetic field (for a review, see 
\cite{Dittrich:2000zu}). For weak magnetic fields, the energy density 
is directly related to the (Euclidean) effective-action density. 
Provided that our present results also hold for $D=4$ spinor QED at 
least qualitatively, we can exclude any further influence of the 
drop-off region, since the magnetic diffusion in this case ($m\gg B$) 
occurs at the order of a Compton wavelength $1/m$. This 
``zero-result'' is supported by considerations within a derivative 
expansion, where the natural expansion parameter is given by the 
Compton wavelength over the length of the field variation (e.g., 
$\simeq 4\times 10^{-12}$ for the PVLAS). Only the constant-field 
result integrated over the size of the magnet including the drop-off 
region need be taken into account. 
 
\section*{Acknowledgement} 
We would like to thank W.~Dittrich for helpful discussions and for 
carefully reading the manuscript. H.G. gratefully acknowledges 
insightful discussions with C.~Schubert. K.L. is indebted to 
H.~Reinhardt for encouragement and support. 
This work has been 
supported in part by the Deutsche Forschungsgemeinschaft under 
contract Gi 328/1-1.

\end{document}